\documentclass[aps,reprint,nofootinbib,nobibnotes,notitlepage,superscriptaddress,twocolumn,prl,
 amsmath,amssymb
]{revtex4-2}

\usepackage[caption=false]{subfig}
\usepackage{braket}
\usepackage{float}
\usepackage{lipsum}
\usepackage{graphicx}% include figure files
\usepackage{dcolumn}% align table columns on decimal point
\usepackage{bm}% bold math
\usepackage{natbib}
\usepackage{hyperref}
\hypersetup{
	colorlinks = true,
    linkcolor = Red,
    urlcolor  = Red,
    citecolor = Red
}
\usepackage{subfig}

\usepackage{microtype}

\usepackage{verbatim}
\usepackage[amssymb]{SIunits}
\usepackage{tabularx}
\usepackage[dvipsnames]{xcolor}
\usepackage{wasysym}
\usepackage[export]{adjustbox}
\usepackage{multirow}
\usepackage{tikz}

\def\be{\begin{equation}}
\def\ee{\end{equation}}
\usepackage{enumitem}

\usepackage{color}
\definecolor{darkgreen}{RGB}{0,120,0}

\newcommand{\av}[1]{\left\langle{#1}\right\rangle} 

\newcommand{\vk}{\vec k}

\def\beq{\begin{eqnarray}}
\def\eeq{\end{eqnarray}}
\let\vec\mathbf

\usepackage{empheq}

\usepackage{booktabs}

\begin{document}

\title{The Limitations of Cosmological Collider Analyses}
%\date{\today}

\author{Oliver~H.\,E.~Philcox}
\email{ohep2@cantab.ac.uk}
\affiliation{Leinweber Institute for Theoretical Physics at Stanford, 382 Via Pueblo, Stanford, CA 94305, USA}
\affiliation{Kavli Institute for Particle Astrophysics and Cosmology, 382 Via Pueblo, Stanford, CA 94305, USA}

\begin{abstract} 
\noindent
Massive fields exchanged during inflation source `cosmological collider' features in the correlators of primordial curvature fluctuations. While their structure is fixed by the symmetries of inflation, the amplitude of such signatures is \textit{a priori} unknown, though bounded by physical principles such as unitarity. We assess the observational viability of the scalar cosmological collider by combining precise numerical predictions derived using \textsc{CosmoFlow} with observational data from \textit{Planck} bispectrum reconstructions. Working in the limit of weak quadratic mixing between the Goldstone and scalar sectors (as assumed by all previous searches), we find that perturbativity restricts the collider signals to be unobservably small, except for double- and triple-exchange configurations at low masses. Larger bispectra can be generated by increasing the mixing to non-perturbative (but still unitary) values: however, this distorts the signal, damping its amplitude and shifting the collider oscillations. Furthermore, strong mixing breaks the factorization of the bispectrum into a free amplitude and a fixed shape, and complicates both theoretical predictions and observational searches. Despite the challenges, we argue that strong mixing analyses are \textit{necessary} to obtain meaningful constraints on the scalar cosmological collider from current and upcoming datasets.
\end{abstract}

\maketitle

\newcommand{\ccs}{c_s}%\textcolor{purple}{c_s}}
\newcommand{\ccspow}[1]{c_s^{#1}}%\textcolor{purple}{c_s^{\textcolor{black}{#1}}}}
\newcommand{\m}{m}%\textcolor{purple}{m}}
\newcommand{\meff}{m_{\rm eff}}%\textcolor{purple}{m_{\rm eff}}}
\newcommand{\rrho}{\rho}%\textcolor{red}{\rho}}
\newcommand{\kkappa}{\kappa}%\textcolor{darkgreen}{\kappa}}
\newcommand{\llambda}{\lambda}%\textcolor{blue}{\lambda}}
\newcommand{\aalpha}{\alpha}%\textcolor{darkgreen}{\alpha}}
\newcommand{\DDelta}{\Delta\rho}%\textcolor{darkgreen}{\Delta\rho}}
\newcommand{\mmu}{\mu}%\textcolor{darkgreen}{\mu}}
\newcommand{\ccthree}{\tilde{c}_3}%\textcolor{blue}{\tilde{c}_3}}

\noindent 
Inflation provides a natural laboratory for studying high-energy particle physics. By combing late Universe correlators for subtle remnants of primordial interactions, we hope to reconstruct the field content and dynamics of the early Universe, and thus constrain the effective inflationary Lagrangian. Since its inception \citep{Chen:2009zp}, this program, often known as `cosmological collider physics' \citep{Arkani-Hamed:2015bza}, has been the subject of intense study across both high-energy theory and observational cosmology. 

Central to this effort are the non-Gaussian correlation functions of the primordial curvature fluctuation, $\zeta$. As shown in \citep{Chen:2009zp,Arkani-Hamed:2015bza,Lee:2016vti}, the exchange of a mass-$m$ scalar field during inflation induces a three-point function with the following squeezed limit (for $k_1\ll k_2\approx k_3$):
\beq\label{eq: collider-signal}
    \av{\zeta(\vk_1)\zeta(\vk_2)\zeta(\vk_3)}' \sim \frac{1}{(k_1k_3)^3}\left(\frac{k_1}{k_3}\right)^{3/2\pm i\sqrt{m^2/H^2-9/4}},
\eeq
%dropping a momentum conserving delta function and denoting the Hubble scale by $H$), 
featuring oscillations whose frequency depends only on the mass. Since this signal can be predicted from symmetry-arguments alone \citep{Arkani-Hamed:2015bza}, its detection would be clear evidence of multi-field inflation.
%Detecting such a signature would be a clear indication of massive fields in inflation, detection of this signature would transform our understanding of the high-energy Universe.

Since primordial curvature fluctuations provide the initial conditions for the evolution of cosmic structure, one can probe collider oscillations using late-time datasets such as the cosmic microwave background anisotropies (CMB). Practically, this combines precise predictions for the primordial correlators (across all kinematic regimes) with non-Gaussianity estimators, performing targeted searches for particular models of interest. Previous studies have applied this to CMB data from \textit{Planck} \citep{Kumar:2026ogn,Kumar:2026dih,Suman:2025tpv,Suman:2025vuf,Salcedo:2026sdn,Philcox:2025bbo,Sohn:2024xzd,Philcox:2026bfa,Philcox:2026njr,Cassem:2026ygh}, as well as galaxy clustering from the Sloan Digital Sky Survey \citep{Cabass:2024wob,Cabass:2022oap,Green:2023uyz,Green:2026yev}, though no detections have yet arisen.

Previous studies have operated in the \textit{weak mixing} limit (often implicitly), wherein the quadratic interactions between the Goldstone mode driving inflation and the exchange field can be treated perturbatively. In this regime, the correlation function factorizes into two pieces: a dimensionless shape, $S_\zeta$ (encoding the structure of the interaction vertex and the properties of the exchange field), and an amplitude, $f_{\rm NL}$ (set by the coupling coefficients): 
\beq\label{eq: fnl-shape}
    \frac{5\av{\zeta(\vk_1)\zeta(\vk_2)\zeta(\vk_3)}'}{18\left[(P_\zeta(k_1)P_\zeta(k_2)P_\zeta(k_3)\right]^{2/3}} = f_{\rm NL}\times S(k_1,k_2,k_3),
\eeq
where $P_\zeta$ is the power spectrum of $\zeta$, and we fix $S(k_*,k_*,k_*)=1$ at $k_* = 0.05\,\mathrm{Mpc}^{-1}$. With this decomposition, searches for cosmological collider physics are reduced to one-dimensional template fits for each shape of interest \citep[e.g.,][]{Sohn:2024xzd,Cabass:2024wob}.

% Crucially, the shape encodes the macrophysics (\textit{i.e.}\ the interaction vertex and properties of the exchange field), while the amplitude depends only on the coupling constants of the theory.
% Fixing the macrophysics (\textit{i.e.}\ picking some shape of interest), we can then probe the cosmological collider signal by searching our datasets for the corresponding value of $f_{\rm NL}$.
% encoding the microphysical coupling constants, 

Though powerful, the weak mixing assumption has its limitations. As discussed in \citep[e.g.,][]{Lee:2016vti}, the size of the collider signal scales with the mixing strength; coupled with the perturbativity restrictions on the cubic inflationary couplings \citep[e.g.,][]{Pinol:2023oux,Kumar:2026dih}, this leads to an upper bound on the $f_{\rm NL}$ amplitude. As such, a set of observations can only meaningfully constrain the collider model if $\mathrm{max}(|f_{\rm NL}|)$ exceeds the experimental sensitivity $\sigma(f_{\rm NL})$. %A key goal of this \textit{Letter} will be to quantify this hierarchy for scalar-exchange models.

Larger signals can be generated by extending to the \textit{strong mixing} regime, which relaxes the bounds on quadratic and cubic couplings \citep{Werth:2023pfl,Pinol:2023oux,Jazayeri:2023xcj}. This is non-trivial, since most methods for computing the inflationary signal assume weak mixing (e.g., the cosmological bootstrap \citep{Arkani-Hamed:2018kmz}). Moreover, strong mixing distorts the bispectrum shape \citep{Pinol:2023oux}, breaking the factorizability of \eqref{eq: fnl-shape} (effectively sending  $S\to S(f_{\rm NL})$). This complicates CMB analyses, prohibiting the use of a single template to represent a wide variety of microphysical models (see Fig.\,\ref{fig: shapes}).

In this \textit{Letter}, we assess the detectability of cosmological collider signals in the weak and strong mixing regimes. We focus on the bispectrum reconstructed from \textit{Planck} temperature and polarization data (noting that upcoming experiments will improve constraints by less than an order of magnitude \citep{SimonsObservatory:2018koc,Cabass:2022epm}), and specialize to the exchange of (principal series) scalar fields with general sound-speeds. Extension to other scenarios, such as higher-spin and chemical potentials, is left for future work.

%- how do we find signatures?
%cmb \& lss. note need the *full* signal, not just this one
%- a central principle: scaling symmetry
%- can these signals be detected? historically we just let fNL be whatever. but it can't be arbitrarily large -- there are theory bounds. also, at large signals, templates distort, need better ways to detect them.
%- goal of this letter: check the detectability in optimistic scenarios for the simplest models: scalar exchange (up to any order). use careful theory calculations including strong-mixing. 

\section{Setup}
\noindent We start from the cubic Lagrangian for the normalized Goldstone mode $\pi_c$ and a scalar field $\sigma$ \citep[e.g.,][]{Senatore:2010wk,Pinol:2023oux}:
\beq\label{eq: lagrangian-cosmoflow}
    \mathcal{L} / a^3 &=& \textcolor{purple}{\frac{1}{2} \left[ \dot{\pi}_c^2 - c_s^2\frac{(\partial_i \pi_c)^2}{a^2} \right]}+ \textcolor{purple}{\frac{1}{2} \left[ \dot{\sigma}^2 - \frac{(\partial_i \sigma)^2}{a^2} - m^2 \sigma^2 \right]}\nonumber\\
    &&\,+\,\textcolor{red}{\rho\dot{\pi}_c \sigma}-\textcolor{blue}{\lambda (\partial_\mu \pi_c)^2 \dot{\pi}_c} -\textcolor{blue}{\Delta\lambda \dot{\pi}_c^3}\nonumber\\
    &&\,-\,\textcolor{darkgreen}{\frac{\kappa}{2}(\partial_\mu \pi_c)^2\sigma}-\textcolor{darkgreen}{\frac{\Delta\kappa}{2}\dot{\pi}_c^2 \sigma} -\textcolor{darkgreen}{\frac{\alpha}{2} \dot{\pi}_c \sigma^2 -\mu\sigma^3}
\eeq
where $\ccs\equiv  c_\pi/c_\sigma$ is the relative sound-speed (setting $c_\sigma=1$) and $\m$ is the scalar mass. This contains four types of term, indicated by color: \textcolor{purple}{kinetic contributions} (controlling the background dynamics), \textcolor{red}{quadratic mixing}, \textcolor{blue}{self-interactions}, and \textcolor{darkgreen}{cubic mixings}. The coupling amplitudes can be written:
\beq\label{eq: coeff-def}
    &&\lambda = \frac{\ccspow{3/2}}{2f_\pi^2}(1-\ccspow{2}), \quad \Delta\lambda = -\frac{\ccspow{3/2}}{3f_\pi^2}(\ccspow{-2}-1)\ccthree,\\\nonumber
    &&\kappa = \frac{\rrho}{f_\pi^2}\ccspow{3/2}, \,\,\quad\qquad \Delta\kappa = \frac{\DDelta}{f_\pi^2}\ccspow{3/2},
\eeq
where $f_\pi \equiv (2M_{\rm Pl}^2|\dot{H}|\ccs)^{1/4}$ is the symmetry-breaking scale, and $\llambda$ and $\kkappa$ are fixed by the quadratic theory (due to the non-linear realization of boosts). Here, the model is parametrized by three quadratic variables: $\{\rrho, \ccs, \m\}$, and four cubic amplitudes: $\{\ccthree, \DDelta, \aalpha, \mmu\}$.
%(noting that $\ccthree=0$ and $\DDelta=0$ lead to boost-invariant couplings, $\propto(\partial_\mu\pi_c)^2$). 

From \eqref{eq: lagrangian-cosmoflow}, we can compute the three-point function of primordial curvature (related to the Goldstone via $\zeta = -(H\ccspow{3/2}/f_\pi^2)\pi_c$). At tree-level, this contains two types of contribution:
\begin{itemize}
    \item \textcolor{blue}{Self-interactions} sourced by $(\partial_\mu\pi_c)^2\dot\pi_c$ and $\dot\pi_c^3$.
    \item \textcolor{darkgreen}{Exchange interactions} sourced by $(\partial_\mu\pi_c)^2\sigma$ and $\dot\pi_c^2\sigma$ (single-exchange), $\dot\pi_c\sigma^2$ (double-exchange) and $\mu\sigma^3$ (triple-exchange).
\end{itemize}
In the weak mixing regime, defined by $\rrho\ll H$, the latter bispectra scale as $\rrho\DDelta$, $\rrho^2\aalpha$, and $\rrho^3\mmu$
%\ccspow{-2}, (1-\ccspow{-2})\ccthree$, $\ccspow{-2}\rrho^2$, $\rrho\DDelta$, $\rrho^2\aalpha$, and $\rrho^3\mmu$ 
respectively. % (with the first and third amplitudes fixed by the quadratic theory). 
For $\rrho\gtrsim H$, these scalings break down since the mixing modifies the propagation of $\pi_c$.

The couplings in \eqref{eq: lagrangian-cosmoflow} are bound by various theoretical considerations including perturbativity and unitarity \citep{Jazayeri:2022kjy,Pinol:2023oux}. For example, restricting the strong coupling scale to be less than $H$ implies $c_s\gtrsim 0.01$ and $|\tilde{c}_3|(\ccspow{-2}-1)\lesssim 10^4$ \citep{Cheung:2007st,Baumann:2011su,Senatore:2009gt}. Assuming weak mixing, perturbativity gives the (order-of-magnitude) bounds
\beq\label{eq: bounds-weak}
    && \frac{\rrho}{H}\lesssim \mathrm{min}\left(\ccspow{-1/2}, \frac{\m}{H}\right), \quad \left|\frac{\DDelta}{H}\right|\lesssim \frac{\ccspow{-1/2}}{2\pi\Delta_\zeta},\\\nonumber
    && |\aalpha| \lesssim \ccspow{1/2}, \qquad\qquad\qquad\,\, \left|\frac{\mmu}{H}\right|\lesssim 1. 
\eeq
In the strong mixing scenario, the perturbativity of cubic operators (coupled with strong coupling analyses for $\rrho$) implies
\beq\label{eq: bounds-strong}
    &&\frac{\rrho}{H} \lesssim \frac{\ccs\kappa^{1/2}}{\Delta_\zeta}, \,\,\,\,\quad\qquad \left|\frac{\DDelta}{H}\right|\lesssim \frac{\ccspow{-1/4}}{2\pi \Delta_\zeta}\left(\frac{\rrho}{H}\right)^{3/4},\\\nonumber
    && |\aalpha|\lesssim \ccspow{1/4}\left(\frac{\rrho}{H}\right)^{3/4}, \quad \left|\frac{\mmu}{H}\right|\lesssim \ccspow{-3/4}\left(\frac{\rrho}{H}\right)^{3/4}
\eeq
where $\kappa \equiv 2\Gamma(5/4)^2/\pi^3$.\footnote{\citep{Kumar:2026dih} imposed the tighter bound $|\alpha| \lesssim 10^{-4}$, assuming $\dot\pi_c^2\sigma$ and $\dot\pi_c\sigma^2$ to arise from a common source.} These will be used to obtain maximum $f_{\rm NL}$ amplitudes for each interaction channel.

\section{Methods}
\noindent We compute the two- and three-point functions of $\pi_c$ using the \textsc{CosmoFlow} code \citep{Werth:2024aui}.\footnote{The two-point function is used to convert from $\pi_c$ to $\zeta$ \citep[cf.][]{cosmoflow1}. For $\rrho\gtrsim H$, $\av{\pi_c^2}$ is enhanced which suppresses $f_{\rm NL}$.} This obtains the correlators by solving differential equations numerically, starting from Bunch-Davies initial conditions \citep{Werth:2023pfl,Pinol:2023oux} (see \citep{An:2017hlx,Kumar:2026ogn,Kumar:2026dih} for an alternative approach). Our implementation is described in detail in \citep{cosmoflow1} and has been carefully verified against the analytic self-interaction results \citep[e.g.,][]{Senatore:2009gt} and the single-exchange bootstrap predictions \citep{Pimentel:2022fsc,Jazayeri:2022kjy,Wang:2022eop}. Working at unit cubic couplings, we compute spectra for each of the six types of interactions across $\mathcal{O}(100)$ kinematic configurations, spanning a wide range of exchange masses, sound-speeds, and mixing coefficients. The full tree-level cosmological-collider bispectrum is assembled as follows:\footnote{An interactive visualization of this is available at \href{https://oliverphilcox.github.io/cosmological-collider-bispectra/}{oliverphilcox.github.io/cosmological-collider-bispectra}.}
\beq\label{eq: explicit-model}
    \av{\zeta(\vk_1)\zeta(\vk_2)\zeta(\vk_3)}'_{\rm true} &=& \sum_{i=1}^6 A_i(\rrho,\ccthree,\DDelta,\aalpha,\mmu)\\\nonumber
    &&\quad\,\times\,b_i(k_1,k_2,k_3;\m,\ccs,\rrho),
\eeq
where $\{A_i\} \equiv \{\ccspow{-2},(1-\ccspow{-2})\tilde{c}_3,\rrho^2,\rrho\DDelta,\rrho^2\aalpha,\rrho^3\mmu\}$. For $\rrho\ll H$, the $b_i$ templates become independent of $\rho$, such that the signal splits into an $f_{\rm NL}$ amplitude (encoding the couplings) and a shape (depending on $\m$ and $\ccs$). We further consider a \textit{scaled} bispectrum model, where $\rho$ is treated perturbatively:
\beq\label{eq: explicit-model-scaling}
    \av{\zeta(\vk_1)\zeta(\vk_2)\zeta(\vk_3)}'_{\rm scaling} &=& \sum_{i=1}^6 A_i(\rrho/\rrho_*,\ccthree,\DDelta,\aalpha,\mmu)\\\nonumber
    &&\quad\,\times\,b_i(k_1,k_2,k_3;\m,\ccs,\rrho_\star)
\eeq
where $\rrho_\star\ll H$. This is assumed in most theory calculations \citep[e.g.,][]{Lee:2016vti,Pimentel:2022fsc,Jazayeri:2022kjy,Arkani-Hamed:2018kmz,Qin:2022fbv} (though see \citep{Pinol:2023oux,Kumar:2026dih,Kumar:2026ogn,An:2017hlx}) and previous scalar-exchange analyses \citep{Kumar:2026ogn,Kumar:2026dih,Suman:2025tpv,Suman:2025vuf,Sohn:2024xzd,Philcox:2026njr,Cabass:2024wob}.

To compare the theoretical models of \eqref{eq: explicit-model}\,\&\,\eqref{eq: explicit-model-scaling} to data, we use the \textit{Planck} binned shape function reconstructions discussed in \citep{Philcox:2026njr} (fixing $\Delta^2_\zeta\propto \av{\zeta^2}$ to the \textit{Planck} best-fit \citep{2020A&A...641A...6P}). These encode $\av{\zeta^3}/\av{\zeta^2}^2$ in a set of $k_1/k_3$ and $k_2/k_3$ bins, and facilitate Fisher forecasts (used to obtain $\sigma(f_{\rm NL})$) as well as detailed searches for collider signatures \citep{cosmoflow1}.

\begin{figure}
    \centering
    \includegraphics[width=0.9\linewidth]{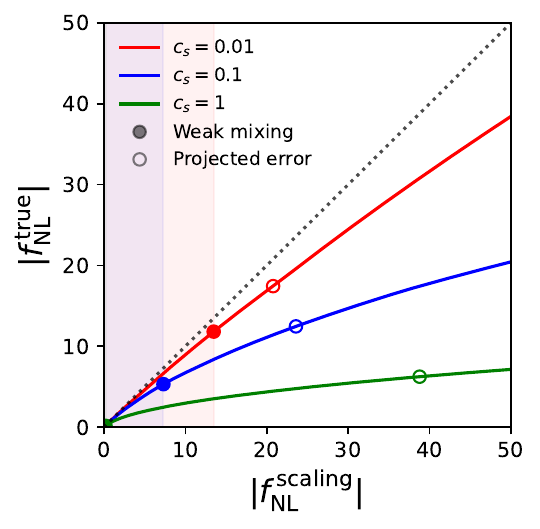}
    \caption{\textbf{Impact of mixing on the bispectrum amplitude.} We compare $f_{\rm NL}$ amplitudes from the scaling regime, assumed in most analyses, to direct computation using \textsc{CosmoFlow}. 
    Results are shown for the $(\partial_\mu\pi_c)^2\sigma$ vertex with mass $\m=3H/\sqrt{2}$ and three choices of sound-speed $\ccs$, with the quadratic mixing set by $\rho\propto (f_{\rm NL}^{\rm scaling})^{1/2}$. Filled circles and shaded regions represent the maximum $f^{\rm scaling}_{\rm NL}$ allowed by weak mixing (cf.\,\ref{eq: bounds-weak}), while open circles give the \textit{Planck} $1\sigma$ errors. For weak mixing, the scaling approximation holds, but yields amplitudes too small to detect. Larger signals require strong mixing, for which $f_{\rm NL}^{\rm true}$ scales slower than expected with $\rho$.}\label{fig: fnl-scaling}
\end{figure}

\section{Results}

\noindent Fig.\,\ref{fig: fnl-scaling} compares the amplitudes of the true and scaling bispectra (\ref{eq: explicit-model}\,\&\,\ref{eq: explicit-model-scaling}) for the single-exchange interaction induced by $(\partial_\mu\pi)^2\sigma$, whose size is controlled by $\rho/H$ (cf.\,\ref{eq: coeff-def}).
%to the weak mixing prediction, defined by $f_{\rm NL}^{\rm scaling}(\rrho) \equiv (\rrho/\rrho_*)^2f_{\rm NL}^{\rm true}(\rrho_*)$ for $\rrho_\star\ll H$ (obtained by combining \ref{eq: fnl-shape}\,\&\,\ref{eq: explicit-model-scaling}). 
%First, we consider the form of the exchange bispectra in the weak and strong mixing regime. For definitiveness, we focus on a single interaction (from $(\partial_\mu\pi)^2\sigma$ at $\sqrt{m^2/H^2-9/4}=3/2$), though our conclusions extend also to other scenarios. 
%which sets both the amplitude of the mixing vertex \eqref{eq: coeff-def} and the mixing between $\pi_c$ and $\sigma$, 
%with $f_{\rm NL}^{\rm scaling}\propto \rrho^2$. 
%In the limit of a small signal or a low sound-speed,\footnote{Single-field analyses constrain $\ccs\gtrsim 0.02$ \citep{Planck:2019kim}.} the bispectrum amplitude scales as expected with $f_{\rm NL}^{\rm true}\approx f_{\rm NL}^{\rm scaling}\propto \rrho^2$. 
In the limit of a small signal or low $\ccs$,\footnote{The enhancement at low $\ccs$ is specific to the $(\partial_\mu\pi_c)^2\sigma$ vertex \citep{Jazayeri:2023xcj}.} the weak mixing regime applies, whereupon $f_{\rm NL}^{\rm true}\approx f_{\rm NL}^{\rm scaling}\propto \rho^2$. At larger $\rrho$, however, we enter the strong mixing regime, for which the signal is suppressed due to the rapid interconversion of $\pi_c$ and $\sigma$ (which reduces $\zeta/\pi_c\propto f_\pi^{-2}$ by enhancing $\av{\pi_c^2}$ \citep{Pinol:2023oux}). 
From \eqref{eq: bounds-weak}, the transition from weak to strong mixing occurs at $f_{\rm NL}^{\rm scaling} \sim \mathcal{O}(10)$, which is a factor of a few \textit{lower} than the forecasted bounds on $f_{\rm NL}^{\rm scaling}$ from \textit{Planck}. We conclude that \textit{the collider amplitude is $\mathcal{O}(1)$ suppressed by mixing in observationally-relevant regimes.}

\begin{figure*}
    \centering
    \includegraphics[width=0.85\linewidth]{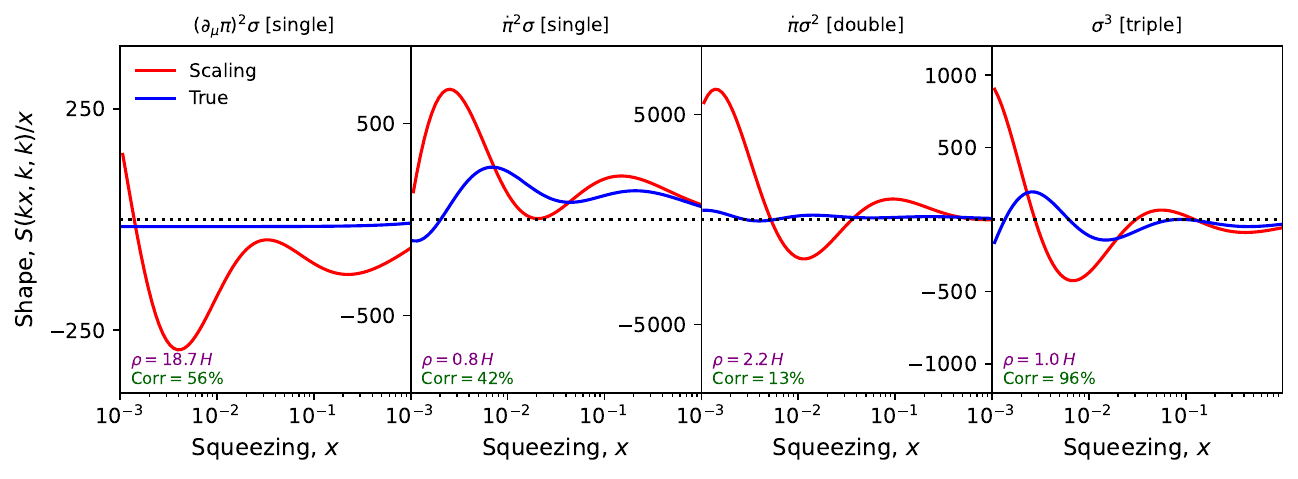}
    \caption{\textbf{Impact of mixing on the bispectrum shape.} We plot the isosceles bispectrum as a function of the triangle squeezing for the four tree-level exchange bispectra, fixing $\m=3H/\sqrt{2}$ and $\ccs=1$. Red lines show the scaling bispectra \eqref{eq: explicit-model-scaling} (assumed in contemporary data analyses), with amplitudes set by the \textit{Planck} 95\% bound. Blue lines show the corresponding true bispectra \eqref{eq: explicit-model}, fixing the quadratic mixing to the smallest value consistent with perturbativity (with numerical values shown in purple). Even in this best-case scenario, the true and assumed shapes differ significantly (with CMB cosines shown in green, projecting out the self-interaction templates from \citep[App.\,B]{Senatore:2009gt}), suggesting that strong mixing is important to account for in data analysis pipelines.}
    \label{fig: shapes}
\end{figure*}

\begin{figure*}
    \centering
    \includegraphics[width=0.85\linewidth]{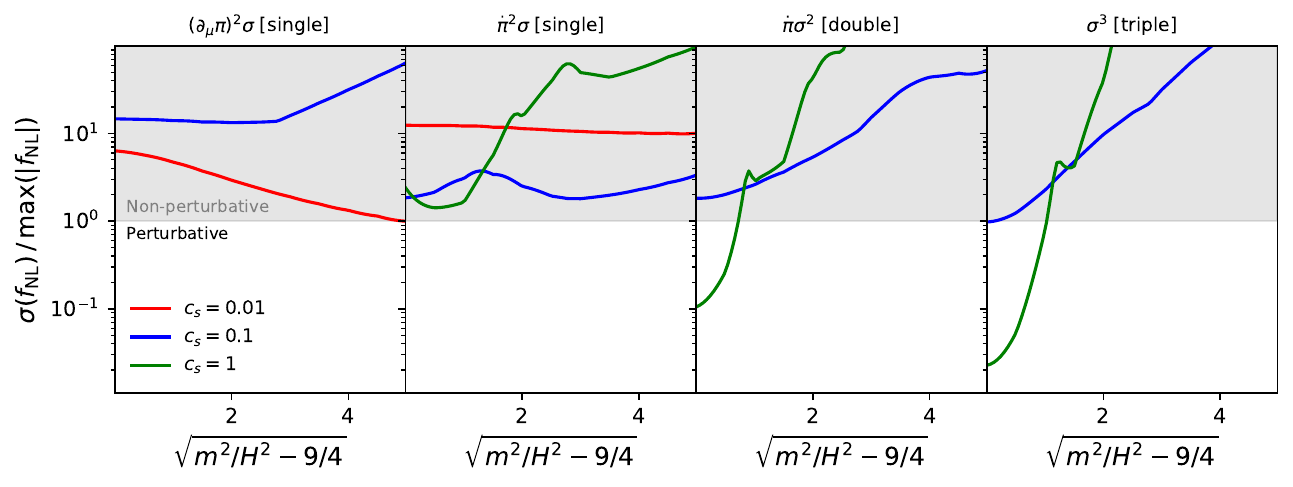}
    \caption{\textbf{Detectability of the weakly-mixed collider.} We compare the \textit{Planck} errorbar on $f_{\rm NL}$ to the maximum amplitude allowed by perturbativity \eqref{eq: bounds-weak}, working in the scaling limit (matching previous studies, and exact only for weak mixing) and projecting out the single-field templates from \citep[App.\,B]{Senatore:2009gt}.
    Ratios above unity indicate scenarios that cannot be meaningfully constrained using current datasets. Though double- and triple-exchange interactions are detectable for $\m\lesssim 2H$ and $\ccs\approx 1$, all other configurations have $\sigma(f_{\rm NL})>\mathrm{max}(|f_{\rm NL}|)$, implying that current observational constraints are broadly inconsistent with perturbativity bounds.
    %will require extension to the strong mixing regime (Fig.\,\ref{fig: fnl-strong}) \oliver{mention that most are ruled out?}.}
    }
    \label{fig: fnl-weak}
\end{figure*}

Mixing also distorts the shape of the inflationary bispectrum. In Fig.\,\ref{fig: shapes}, we compare the scaled and full correlators across all scalar-exchange diagrams, choosing the coupling coefficients such that each scaling solution saturates the 95\% upper-bound from \textit{Planck} and the quadratic mixing takes the smallest value consistent with the perturbativity bounds of \eqref{eq: bounds-strong}.\footnote{We fix $\ccs=1$ to maximize the detectability of the double- and triple-exchange signals (see Fig.\,\ref{fig: fnl-weak}).} To reach the detection thresholds, $\rho/H\gtrsim\mathcal{O}(1)$ is required: this leads to strong distortions of the bispectrum shape. In particular, the collider oscillations shift to higher frequency with damped amplitudes, or decay entirely (for $(\partial_\mu\pi_c)^2\sigma$). This matches the results of \citep{Pinol:2023oux}, who note that the squeezed limit \eqref{eq: collider-signal} becomes controlled by the \textit{effective} mass, $\meff=\sqrt{\m^2+\rrho^2}$, rather than $m$ (with oscillations damped as $e^{-\pi m_{\rm eff}}$). The large distortions, whose self-interaction-deprojected correlations reach as low as $13\%$) indicate that \textit{the scaling templates used in contemporary searches differ considerably from the true underlying spectra}, motivating a full analysis of the collider paradigm.

Next, we assess the detectability of the cosmological collider, starting with the weak mixing limit (where the scaling limit applies). Fig.\,\ref{fig: fnl-weak} compares the maximum amplitudes allowed by the theoretical bounds of \eqref{eq: bounds-weak} to the $1\sigma$ errorbars from \textit{Planck}. To isolate the collider phenomenology of interest, we modify the templates by subtracting the overlap with the equilateral and orthogonal shapes following \citep{Suman:2025tpv,Suman:2025vuf}.
%\footnote{Note that this yields tighter constraints than if one marginalizes over equilateral and orthogonal shapes.} 
For single exchange, $\sigma(f_{\rm NL})\gtrsim \mathrm{max}(|f_{\rm NL}|)$ for all masses and sound-speeds, indicating that any collider signals are hidden by noise.\footnote{Given the order-of-magnitude nature of the perturbativity bounds, there is some hope for detecting the $\dot\pi_c^2\sigma$ signal at $\ccs\approx 1$.} Double- and triple-exchange interactions can yield detectable amplitudes for $m\lesssim 2H$ \citep[cf.][]{Kumar:2026dih,Kumar:2026ogn}, though we caution that the underlying bispectra become suppressed as we approach the $\rho=H$ weak mixing bound (cf.\,Fig.\,\ref{fig: shapes}). 
We conclude that \textit{much of the parameter space probed by scaling analyses is already ruled out by perturbativity constraints}.

\begin{figure*}
    \centering
    \includegraphics[width=0.85\linewidth]{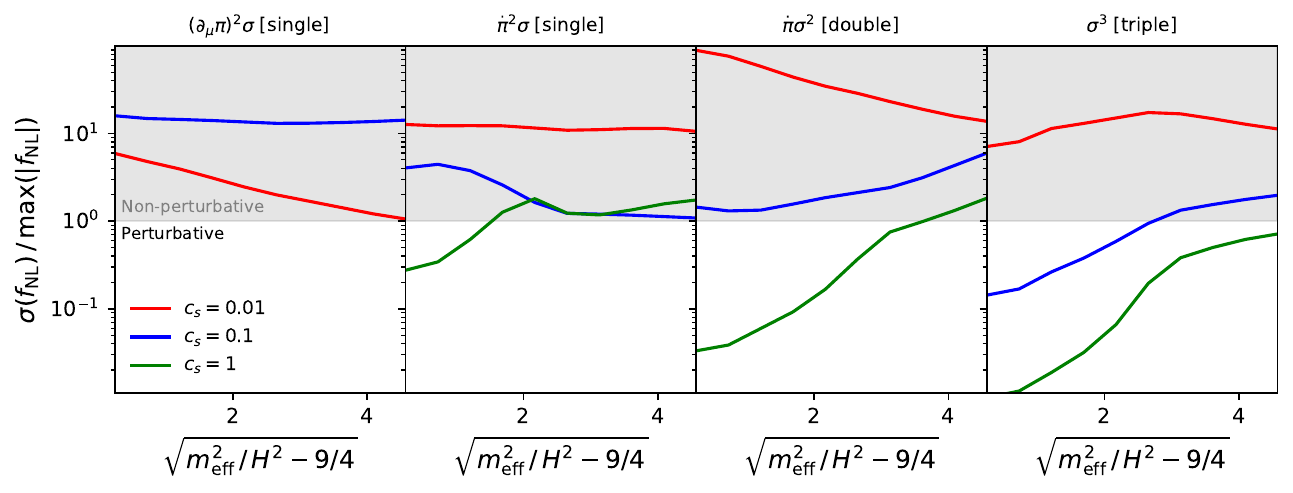}
    \caption{\textbf{Detectability of the strongly-mixed collider.} As Fig.\,\ref{fig: fnl-weak}, but extending to $\rho\gtrsim H$. In each case, we show the model with maximal signal-to-noise across all parameter sets with a given effective mass, $\meff = \sqrt{\m^2+\rho^2}$ (which sets the squeezed-limit behavior). While the collider signal sourced by $(\partial_\mu\pi_c)^2\sigma$ remains too small to observe, the other shapes become detectable across a wide range of masses and sound speeds, motivating the observational searches carried out in \citep{cosmoflow1}.}
    \label{fig: fnl-strong}
\end{figure*}

Analogous results for the strongly-mixed collider are shown in Fig.\,\ref{fig: fnl-strong}. Here, we maximize the detectability by fixing the cubic couplings to the largest values allowed by perturbativity \eqref{eq: bounds-strong}, and optimizing over all masses and quadratic mixings at a given $m_{\rm eff}=\sqrt{m^2+\rho^2}$.\footnote{Since collider oscillations depend on the \textit{effective} mass, we include complementary series fields ($m<3H/2$) in this analysis.}
%As noted above, the strong mixing shapes gain explicit (non-scaling) dependence on the mixing amplitude $\rrho$; to account for this, we fix the cubic coupling amplitudes to the maximal values from \eqref{eq: bounds-strong}, and maximizing the detectability over $\rrho$, giving results as a function of the effective mass $m_{\rm eff}$, which controls the phenomenology.
% , we forecast the detectability of the strong mixing collider, obtained through Fisher analyses of \eqref{eq: explicit-model}, fixing the cubic coupling amplitudes to the maximal values from \eqref{eq: bounds-strong} and maximizing the detectability over $\rrho$. 
As for weak mixing, both the $(\partial_\mu\pi_c)^2\sigma$ interaction and low-$c_s$ models are undetectable with current datasets: large signals require $\rrho\gg H$ and thus $\meff\gg H$, which damps out the target collider oscillations. For the other interactions, we find clear signals for all $\meff$ (primarily from $\rrho\gtrsim H$) across a wide range of $\ccs$; \textit{this represents a key area of opportunity for future observational studies of the cosmological collider.} %Since these have $\rrho/H\gtrsim \mathcal{O}(1)$, their prediction requires a full numerical computation, extending beyond the scaling limits. 
%The conclusion is clear: \textit{theory permits a considerable collider parameter space in the (semi-)strong mixing regime}.

\section{Discussion}
\noindent Previous searches for cosmological collider physics have assumed that the mixing between the new field and the Goldstone mode can be treated perturbatively \citep{Cabass:2024wob,Sohn:2024xzd,Suman:2025tpv,Suman:2025vuf,Philcox:2026njr,Kumar:2026ogn,Kumar:2026dih}. Once theoretical constraints on the coupling coefficients are accounted for, this leads to undetectably small signals across most of parameter space (see Fig.\,\ref{fig: fnl-weak}). 
%, this leads to undetectably small non-Gaussianity across most of parameter space, folding in contemporary errorbars from the CMB \citep{Philcox:2026njr} and theoretical constraints on the underlying coupling coefficients. 
Our conclusions match those of \citep{Kumar:2026dih}, which also excluded the double-exchange interaction, due to tighter priors. This implies that many of the models used in previous studies should be interpreted as \textit{effective} templates rather than true theory-derived signatures \citep[cf.][]{Suman:2025tpv}.

Larger signals can be generated by boosting the quadratic mixing parameter $\rho$. However, this comes at a cost: $f_{\rm NL}$ grows more slowly than the scaling prediction (Fig.\,\ref{fig: fnl-scaling}) and the collider oscillations are both damped and shifted to higher frequencies (Fig.\,\ref{fig: shapes}, matching \citep{Pinol:2023oux,Werth:2023pfl,Jazayeri:2023xcj}). Notably, these effects are relevant even for $\mathcal{O}(1)$ mixing (which is commonly assumed to derive upper bounds on $f_{\rm NL}$ \citep[e.g.,][]{Kumar:2026dih}). As shown in Fig.\,\ref{fig: fnl-strong}, the strongly-mixed collider offers much greater detectability than the weakly-mixed equivalent, with theoretically-allowed regimes for all interactions except $(\partial_\mu\pi_c)^2\sigma$ across a broad range of masses and sound-speeds.

% From the above exercise, we conclude that the $f_{\rm NL}$ values probed by most studies (which implicitly or explicitly assume weak mixing) are inconsistent with the theoretical bounds from \eqref{eq: bounds-weak} across much of the parameter space. Can we do better by extending to strong mixing? Such analyses are considerably more difficult, since the factorization into amplitude and shape is lost, e.g., the amplitude of the triple exchange is set by both $\rrho$ and $\mmu$, where $\rrho$ informs both the shape and the theoretical bound on $\mmu$. 

How do we analyze such models in practice? Outside the weakly-mixed regime, the factorization of the bispectrum into a fixed shape and a free amplitude no longer applies. This complicates traditional template-search analyses and requires non-perturbative computation of the inflationary correlators \citep[e.g.,][]{Werth:2023pfl,An:2017hlx}. Moreover, for $\rrho\gtrsim H$, collider oscillations are induced in the self-interaction bispectra sourced by $(\partial_\mu\pi_c)^2\dot\pi_c$ and $\dot\pi_c^3$ \citep{Pinol:2023oux,Jazayeri:2023xcj}. To fully explore the scalar cosmological collider, we require a \textit{joint} analysis of all tree-level contributions, working non-perturbatively in the quadratic mixing $\rrho$. This is a difficult task, but is made possible by recent developments in CMB analysis and inflationary theory, and will be presented in our upcoming work \citep{cosmoflow1}.

Whilst this \textit{Letter} has considered only a single observational dataset (the \textit{Planck} bispectrum), we expect our conclusions to remain valid at least into the next decade. Due to cosmic variance limitations, upcoming CMB experiments are expected to tighten the bounds on primordial non-Gaussianity only by a factor of a few \citep{CMB-S4:2016ple,SimonsObservatory:2018koc} (based on forecasts for the equilateral shape). Despite the rapid influx of spectroscopic data from DESI and \textit{Euclid}, the outlook for LSS probes appears similar, given the current bounds on equilateral non-Gaussianity and the forecasts for next-generation experiments \citep{Cabass:2022epm,Chudaykin:2025vdh}.

Finally, we note that our analysis has restricted to the simplest type of collider non-Gaussianity: tree-level scalar field exchange. Larger signals can be generated from other models, featuring chemical potentials \citep[e.g.,][]{Bodas:2025vpb,Kumar:2026ogn,Kumar:2026dih}, higher spin fields \citep[e.g.,][]{Tong:2022cdz,Sohn:2024xzd}, loops \citep[e.g.,][]{You:2026xoq}, curvatons \citep{Cassem:2026ygh}, strongly-coupled sectors \citep{Pimentel:2025rds,Jiang:2025mlm}, primordial features \citep[e.g.,][]{Chen:2022vzh} and beyond. Moreover, large signals may be present in higher-order correlators, such as the trispectrum \citep[e.g.,][]{Philcox:2025wts}. Much remains to be explored.

\vspace{1.0em}
\acknowledgments
{\footnotesize
\begingroup
\hypersetup{hidelinks}
\noindent 
We thank Soubhik Kumar, Petar Suman and Denis Werth for insightful discussions. OHEP thanks the \href{https://www.flickr.com/photos/198816819@N07/55385954556}{llegions of Xi'An} for guidance. The computations in this work were run at facilities supported by the Scientific Computing Core at the Flatiron Institute, a division of the Simons Foundation. 
\endgroup
\vskip 4pt
}

\appendix

\bibliographystyle{apsrev4-1}
\bibliography{refs}% Produces the bibliography via BibTeX.

\end{document}